\pdfoutput=1
\documentclass[sigchi]{acmart}

\usepackage{booktabs} 
\usepackage[utf8x]{inputenc} 
\usepackage{balance}
\usepackage{titlesec}

\copyrightyear{2019} 
\acmYear{2019} 
\setcopyright{acmlicensed}
\acmConference[CHI 2019]{CHI Conference on Human Factors in Computing Systems Proceedings}{May 4--9, 2019}{Glasgow, Scotland UK}
\acmBooktitle{CHI Conference on Human Factors in Computing Systems Proceedings (CHI 2019), May 4--9, 2019, Glasgow, Scotland UK}
\acmPrice{15.00}
\acmDOI{10.1145/3290605.3300705}
\acmISBN{978-1-4503-5970-2/19/05}

\title[What Makes a Good Conversation?]{What Makes a Good Conversation? Challenges in Designing Truly Conversational Agents}

\begin{document}

\author{Leigh Clark}
\email{leigh.clark@ucd.ie}
\affiliation{University College Dublin, Ireland}

\author{Nadia Pantidi}
\email{konstantia.pantidi@ucc.ie}
\affiliation{University College Cork, Ireland}

\author{Orla Cooney}
\email{orla.cooney@ucdconnect.ie}
\affiliation{University College Dublin, Ireland}

\author{Philip Doyle}
\email{pdoyle@voysis.com}
\affiliation{Voysis Ltd., Ireland}

\author{Diego Garaialde}
\email{diego.garaialde@ucdconnect.ie}
\author{Justin Edwards}
\email{justin.edwards@ucdconnect.ie}
\affiliation{University College Dublin, Ireland}

\author{Brendan Spillane}
\email{brendan.spillane@adaptcentre.ie}
\author{Emer Gilmartin}
\email{gilmare@tcd.ie}
\affiliation{Trinity College Dublin, Ireland}

\author{Christine Murad}
\email{cmurad@taglab.ca}
\affiliation{University of Toronto, Canada}

\author{Cosmin Munteanu}
\email{cosmin@taglab.ca}
\affiliation{University of Toronto Mississauga, Canada}

\author{Vincent Wade}
\email{vincent.wade@adaptcentre.ie}
\affiliation{Trinity College Dublin, Ireland}

\author{Benjamin R. Cowan}
\email{benjamin.cowan@ucd.ie}
\affiliation{University College Dublin, Ireland}

\renewcommand{\shortauthors}{L. Clark et al.}

\begin{abstract}
Conversational agents promise conversational interaction but fail to deliver. Efforts often emulate functional rules from human speech, without considering key characteristics that conversation must encapsulate. Given its potential in supporting long-term human-agent relationships, it is paramount that HCI focuses efforts on delivering this promise. We aim to understand what people value in conversation and how this should manifest in agents. Findings from a series of semi-structured interviews show people make a clear dichotomy between social and functional roles of conversation, emphasising the long-term dynamics of bond and trust along with the importance of context and relationship stage in the types of conversations they have. People fundamentally questioned the need for bond and common ground in agent communication, shifting to more utilitarian definitions of conversational qualities. Drawing on these findings we discuss key challenges for conversational agent design, most notably the need to redefine the design parameters for conversational agent interaction.
\end{abstract}

%
%



\maketitle

\section{Introduction}
The proliferation of Intelligent Personal Assistants (IPAs) (e.g. Siri and Amazon Alexa) and the importance of speech in embodied agent technologies (e.g. robotics and embodied conversational agents (ECAs)) make human-agent conversation a critical topic of study. The IPA market alone is predicted to reach \$9 billion by the early 2020s \cite{businesswire_global_2018}. Speaking to agents currently falls short of conversational dialogue \cite{mctear_conversational_2016,moore_is_2017,porcheron_voice_2018}, instead offering interactions that are constrained and task oriented. Although not necessarily desirable for all interactions, more human-like conversational ability could be important in supporting long-term human-agent interaction. This is especially true in contexts where social interaction and bond are important for delivering particular services (e.g. health, social care or education). 

To address the current deficit in conversational ability, developers have largely focused on imbuing agents with more human-like conversational qualities such as social talk \cite{bickmore_social_2005,spillane_introducing_2017} and humour \cite{devillers_multifaceted_2018}. The general aim is to emulate elements, and thus the qualities, of human conversation \cite{gilmartin_social_2017}.  This approach, while important for generating conversation-like behaviour, risks losing focus on the subjective qualities that make conversation what it is. It also does not take into account the qualities users feel should manifest in conversational agent interactions, and how they should be implemented. Our work contributes to filling this gap by identifying: 1) what characteristics people see as important in conversation, and 2) how these vary when applied to conversations with artificial agents.  

Through semi-structured interviews, we found that people clearly identified social and transactional roles of conversation, with almost universal focus on transactional purposes when discussing agent conversations. Participants emphasized similar characteristics as important in both human and agent-based conversation (e.g. mutual understanding, trustworthiness, active listenership and humour). Yet they were operationalized very differently, being discussed in almost purely functional terms for agent conversations. Participants described how conversational form and purpose vary with friends, strangers and acquaintances, emphasising the importance of conversation in long term relationship development. In reflecting on conversations with agents, participants’ descriptions echoed human conversations with strangers or casual acquaintances. They fundamentally questioned the need to develop relationships with agents, seeing the agent as a tool rather than a dynamic social entity. Our findings are novel in that they contribute important user led insight to the growing debate around developing conversational ability in spoken agents. Based on these findings we emphasise the need to redefine conversational agent design parameters, based on the distinct differences between people’s experiences of human-human and human-agent conversations.

\section{Related Work}
\subsection{What is Conversation?}
Spoken conversation is defined as: "any interactive spoken exchange between two or more people" \cite[p.11]{bruce_role_2002} Human spoken conversation serves many purposes. These are broadly classified as transactional (task-based) or social (interactional) \cite{brown_discourse_1983, eggins_analysing_2005,schneider_small_1988}. Transactional conversation pursues a practical goal, often fulfilled during the course of one interaction. In these types of exchanges, both interlocutors know what the goal of the dialog is. They have different clearly-defined roles, and success is measured by the achievement of the transaction’s purpose. The aim of more social conversation is not to complete a task as such, but to build, maintain and strengthen positive relations with one or more interlocutors \cite{dunbar_grooming_1998, malinowski_problem_1994}. Social conversation ranges from small talk and social greetings to longer interactions. Examples include talk between friends, office chat, or brief discussions between strangers. This type of social conversation can help develop common ground \cite{clark_using_1996}, trust and rapport between interlocutors \cite{cheepen_predictability_1988}. Although transactional and social talk serve different purposes, they often overlap in natural conversation \cite{cheepen_predictability_1988}.

Critical to a conversation is the opening of a channel by an interlocutor, with a commitment from the other to engage, each then using the dialogue to co-construct meaning and converge on agreement \cite{dubberly2009modeling}. Particularly in task oriented dialogue, this may lead to the proposition of an action to be completed or a transaction to take place.

\subsection{Conversational Agents}
The recent popularity of IPA devices like Amazon Echo and Google Home has generated considerable interest in the HCI community \cite{munteanu2017designing, porcheron_voice_2018}. Recent emphasis in HCI has been on understanding user experiences of interactions with IPAs. This work shows that people tend to engage in limited and clearly delineated task-based conversations with IPAs. These include checking the weather, setting reminders, calling and messaging, playing music, launching other applications, information seeking, and interacting with Internet of Things (IoT) devices \cite{cowan_what_2017, luger_like_2016, bentley2018understanding, mehrotra2017hey, tsai2018Alexa}. Although these devices promise much through their implied humanness \cite{cowan_what_2017}, they fall short of the reflexive and adaptive interactivity that occurs in most human-human conversation \cite{porcheron_voice_2018}. Instead interactions revolve around brief question/answer or user-instructions/system-confirmation dialogues \cite{porcheron_voice_2018}. Even where elements of social talk are implemented this seldom transcends the constraint of a single question and response. This is a far cry from the elaborative, contextual social talk seen in human conversation \cite{gilmartin_explorations_2018}. The highly functional and utilitarian nature of IPA interaction has led to suggestions that ‘conversation’ is a poor description of the current interaction experience \cite{mctear_conversational_2016, porcheron_voice_2018}. This deficit has motivated a number of efforts to improve the conversational capabilities in such agents \cite{fang_sounding_2017, papaioannou_alana_2017}. Recent work has looked to imbue systems with the ability to engage in social talk \cite{spillane_introducing_2017} and humour \cite{devillers_multifaceted_2018}.

Although the current focus in HCI has been on commercial IPA experience, research on ECAs and social robotics have explored the user experience of more conversational interactions with agents. A number of these agents have been developed for contexts including health \cite{bickmore_response_2010, bickmore_towards_2004}, elderly care \cite{bickmore_acceptance_2005, vardoulakis_designing_2012}, education \cite{saerbeck_expressive_2010}, customer service \cite{bickmore_relational_2001} and workplace \cite{gockley_designing_2005} contexts. ECA research \cite{bickmore_relational_2001} showed that users trusted an agent more when small talk was used and found the interaction more human like, although this varied by the user’s personality type and level of ECA embodiment \cite{bickmore_social_2005}. Similarly, in social robotics, conversational interaction is described as important to long term rapport building and use \cite{tapus_socially_2007}. The accommodation of social conversations is proposed as a critical factor in developing trust in a social agent \cite{looije_persuasive_2010}. Work by Sabelli et al. \cite{sabelli_conversational_2011} found that people were open to speaking about personal matters with a conversational robot, using it as an emotional outlet that reduced feelings of loneliness.

\subsection{Research Aims}
Social aspects of conversation may not be useful in all agent interactions. However, they may be beneficial in situations where the agent needs to build trust or rapport or needs to engage in frequent long-term user interaction. Emulating the structures and rules that govern human conversation \cite{gilmartin_social_2017} help generate conversation-like abilities, but they do not guarantee that the important subjective characteristics of conversation are preserved. Our work aims to identify: 1) what characteristics people see as important in conversation, and 2) how these vary when applied to conversations with artificial agents. Our goal is not to explore conversational properties, which are well known, but to emphasise the innate differences between the two types of conversations. We achieve this by presenting the results of semi-structured interviews on people’s understandings and expectations of conversations with people and agents, and what contexts they expect speech-capable agents to be most suited to in the future.

\section{Method}
\subsection{Participants}
Seventeen participants (M=9, F=8) were recruited from a university community via internal email. In line with best practice, participants were recruited until saturation had occurred. Demographic data collected as part of the interview showed participants had a mean age of 31.1 years (SD=8.58) and consisted of students and staff across a wide range of schools. The majority of participants were native English speakers (64.7\%), with the remaining participants speaking English to a near-native level (35.3\%). Most participants rated their expertise with technology as advanced (41.2\%) or intermediate (35.3\%), with fewer describing themselves as experts (23.5\%). The majority of participants indicated they had previously used voice-based assistants (76.5\%), although of this majority (58.8\%) said they used them very infrequently. Participants reported interacting with Siri more than other assistants (56.3\%), followed by Alexa (31.3\%) and Google Assistant (31.3\%). Participants were provided with a €10 honorarium in exchange for taking part.

\subsection{Procedure and Data Analysis}
After receiving university ethics clearance, interviews were conducted over a period of three weeks, lasting an average of 40 minutes, and audio-recorded with the participants’ consent. The interviews were semi-structured, providing us with the flexibility to adjust questioning based on participant responses. Each interview covered four central topics: 1) Important characteristics, purposes and experiences of conversations with different types of people e.g. friends, acquaintances, strangers; 2) General attitudes towards conversations with agents; 3) Reflection on the important characteristics identified in (1) in the context of agent conversation; 4) Appropriate scenarios where conversation with agents could be used. The participant-led characteristics discussed in (1) were written down by the interviewer and discussed again during topic (3). These characteristics and the contexts in which they were discussed were participant generated rather than guided by the interviewer. Furthermore, participants were not prompted to discuss conversations with agents prior to being asked by the interviewer. Following the interviews, participants completed a brief online questionnaire to gather demographics. They were then debriefed and thanked for participating. 

The audio recordings were transcribed and analysed in a systematic qualitative interpretation using Inductive Thematic Analysis \cite{braun_using_2006}. Initial coding was conducted by two researchers working independently. We began with an inductive approach, then grouped themes under central topics, informed by the interview guide. Once all data was coded and a set of initial themes was considered, a data session with additional researchers took place. In this session, the researchers closely reviewed the coding and preliminary themes from the data. These themes were further refined by the initial data coders. The transcribed quotes presented in the findings below are representative of the themes they discuss. This follows the same practise in similar qualitative HCI research \cite{cowan_what_2017, luger_like_2016}. All researchers had a background in HCI, with experience in conducting qualitative data analyses. In the following section we present the findings from our analysis. 

\section{Findings}
\subsection{Purposes of Conversation}
Echoing the literature \cite{cheepen_predictability_1988}, two broad purposes for conversation were clearly shown in the data. These reflected social and transactional goals at the core of most discourse. 

\subsubsection{Social Purposes}
The desire to socialise with others was commonly discussed as a principle aim of conversation. Conversation was seen a way of establishing, maintaining and building social bond. Our participants felt conversation was imperative for getting to know people, forming social groups and deepening relationships.
\begin{center}
\emph{"You can go a bit deeper into knowing people. I would say talking and having conversation is the biggest part of knowing somebody." [P101]}
\end{center}
Within social conversation more frivolous talk or ‘having the craic’ may be dynamically interspersed with more serious topics of discussion.\
\begin{center}
\emph{"You can have very serious conversations...but then you’re just talking absolute shite…having the craic and there’s so much historical context for that conversation so it’s enjoyable.” [P106]}
\end{center}

\subsubsection{Transactional Purposes}
Juxtaposing the social nature of conversation, participants described times when discourse is more goal-oriented, allowing for people to gather information they need to complete a clearly delineated task or objective. In these, the conversation may shift or end after the speaker feels their goal have been achieved. 
\begin{center}
\emph{"In some conversations you’re just trying to elicit information and there’s a very clear purpose…There’s a very clear short-term objective to the conversation. It’s very brief, very goal-oriented. Once you achieve that you just move on.” [P109]}
\end{center}

\subsection{Attributes of Conversation}
Participants described several key attributes that they value in conversation. Both the purpose of conversation and whom a conversation is with can change the importance of these attributes and the role they play.  A number of these attributes and their purposes align with issues discussed in existing linguistics literature \cite{brown1987politeness,clark_using_1996,goffman_interaction_1967,jaworski2014silence, ziv_social_2010}.

\subsubsection{Mutual Understanding \& Common Ground}
Establishing common ground with others was often mentioned as an integral feature of good conversation. Participants stressed the importance of understanding the intent and meaning behind what other speakers are saying beyond. This may involve getting to know others through questioning.

\begin{center}
\emph{"Sometimes people don't get what you're saying, even though you could be using simple language and talking about simple stuff. It's important that both of you understand what each other is saying, and I don't mean that in terms of like a language barrier.” [P105]}
\end{center}

\begin{center}
\emph{"The questions you ask and the questions people ask you even more so are telling about who that person is, what information is it that they’re looking for. You’re always trying to find a common ground.” [P108]}
\end{center}

As well as providing a mutually understood focus during interaction, a knowledge of others supports attempts to reach a common understanding.

\subsubsection{Trustworthiness}
Trustworthiness was also discussed as key for conversation through being a key foundation for growing common ground and subsequently sustaining long-term relationships.

\begin{center}
\emph{"Trustworthy is one of the major characteristics for me. I wouldn’t really bother having a long conversation with somebody I knew wasn’t trustworthy. Having personal conversations with them would be quite disastrous.” [P103]}
\end{center}

\begin{center}
\emph{"You build up this reservoir of knowledge about this person where one day [you] go ‘you know what? I can really trust this human being.’ When you get to this stage there’s desire to keep them in your life.” [P106]}
\end{center}

Having trust in a partner seems to be a gateway to open the possibility of more personal conversations. Without this trust, such conversations, which are common in long term and deep relationships, may be seen as inappropriate. 

\subsubsection{Active Listening}
Active listening was also an important partner attribute in conversation. Participants described that paying attention, demonstrating engagement and a willingness to participate in conversation was important in a two-way interactive dialogue.

\begin{center}
\emph{"Yeah, we know when they are not following what we are saying so paying attention to somebody who is speaking is very important.” [P103]}
\end{center}

\begin{center}
\emph{"Wanting to continue the conversation - wanting to know what comes next and being interested in what the other person is saying. I guess feeling that it's two way as a conversation… a conversation has to go both ways as opposed to one person participating." [P109]}
\end{center}

Conversational partners need to demonstrate both good listenership and understanding. In additional to verbal cues, this can be helped by nonverbal feedback like eye contact.

\begin{center}
\emph{"I definitely prefer when people look me in the eye. I find it very unsettling if you know you’re speaking to somebody and they can’t make eye contact with you even if it’s just fleeting eye contact.” [P108]}
\end{center}

\subsubsection{Humour}
Humour was also commonly mentioned. Participants proposed that it scaffolds and adds more substance to discussions. They also remarked how humour can be a key conversational driver. 
\begin{center}
\emph{"I think a bit of humour is fundamental to good conversation.” [P112]}
\end{center}
 
\begin{center}
\emph{"Conversations can just be humorous as well. I find there tends to be some kind of substance underlying it otherwise it’s not engaging.” [P111]}
\end{center}

Although, humour may be an important conversational characteristic, P111’s comments also highlight that humour itself needs substance and relevance to the conversation. The comment alludes to humour’s ability to soften serious intentions or deliver substantive messages in conversation, in a way that may save face for the speaker.  

\subsection{Building Relationships with Conversation}
Participants also discussed the differences in conversational needs depending on the types and stages of relationships they have with interlocutors. 

\subsubsection{Conversing with Friends}
When conversing with friends, participants mentioned relying heavily on shared experiences and a history of trust. Because of this trust, our participants regularly conversed with close friends to get advice or alternate perspectives, using them as sounding boards to offload issues and release personal tension. 

\begin{center}
\emph{"I have a friend who I really trust a lot and respect, so I usually make sure I call her and get her opinion about the situation.” [P103]}
\end{center}

\begin{center}
\emph{"You feel there is someone for you. If you are in a problem and if you need a second advice, then they know [you] from maybe years together.” [P114]}
\end{center}

\begin{center}
\emph{"Sometimes when you’re narrating an incident, you’re just doing that to get a load off your mind.” [P103]}
\end{center}

The type of topics covered with close friends were more personal in nature compared to other types of conversational partners (see followin subsection). Such conversations seem built upon a shared repository of memories and experiences, that are co-constructed.

\begin{center}
\emph{"You have intimate conversations where you would be talking about something personal to them…friend’s grandmother was sick last week. Something like that that is intimate, private, discreet.” [P109]}
\end{center}

\begin{center}
\emph{"Your conversation leads to memories, building memories together, and that’s a huge part of having a friendship that you can reference back and say we did this together and we did that together.” [P106]}
\end{center}

With very close friends, participants noted that there may not be a need for constant conversation, as silence may symbolise a level of comfort between two friends. 

\begin{center}
\emph{ "…with my very close friends, I don’t need to talk - it’s not like we talk at every given moment….” [P101]}
\end{center}

\subsubsection{Conversing with Acquaintances and Strangers}
When compared to conversations with close friends, those with strangers and casual acquaintances often had a more superficial and functional purpose. Common ground is formed from shared context (e.g. the workplace) or from shared identity with the interlocutor (e.g. colleagues) and used as the anchor for conversational topics and content.

\begin{center}
\emph{"With acquaintances it’s not as personal...it would be more about the common area where we know each other. If we are colleagues, we are probably talking about a meeting we attended or common acquaintances at work.” [P103]}
\end{center}

\begin{center}
\emph{ "I usually talk about public life, like work but on a very superficial level.” [P104]}
\end{center}
With acquaintances and strangers, topics covered were more superficial and general in nature, focusing on current events, shared contexts or the need to share information.

\begin{center}
\emph{"...I suppose we talk...like surface level stuff again like the weather...or maybe sharing a complaint like if you are waiting for a bus” [P108]. }
\end{center}

The content of these conversations may be more limited compared to friends as participants noted concern as to what topics and behaviour are deemed appropriate. 

\begin{center}
\emph{"...it sometimes has to be more stilted because you can’t assume everyone will find the same thing appropriate. You have to minimise the scope of the conversation.” [P111]}
\end{center}

\begin{center}
\emph{"...the further you get out from your inner circle [of friends] the constraints come in on what you’re going to speak about.” [P106].}
\end{center}

These interactions may be driven by a perceived social norm to initiate or respond to conversational approaches. For instance, not doing so may seem impolite, particularly if initiated by another speaker. Similarly, participants may use conversation to reduce feelings of awkwardness, especially in periods of silence. 

\begin{center}
\emph{"A lot of people are uncomfortable with silence. When you get very far from the core of very close friends, that discomfort you feel with silence grows…With other people I don’t spend a lot of time with, it’s very important that the time is spent talking.” [P101]}
\end{center}

\begin{center}
\emph{"In terms of day-to-day, you know just meeting people on the street, to me it's social expectation...Even saying nothing is saying something.” [P111]}
\end{center}

Engaging in conversation, even if just small talk, can help make others and oneself feel comfortable in these situations. Especially with strangers, small talk and transactional dialogue were clearly the main drivers of conversation. 

\subsubsection{Transition Towards Friendship}
Participants identified conversation as a fundamental tool used to transition towards friendship, allowing them to ‘gauge’ one another. Being able to share vulnerabilities and personal information was seen as an important step towards developing mutual trust and bond with others.

\begin{center}
\emph{"It’s [conversation] how you gauge a connection and a forge a pathway where you can understand what is going on with this person. I think bonds can be created with much conversation for sure.” [P106]}
\end{center}

\begin{center}
\emph{"I think sharing a vulnerability no matter how small or even acknowledging a sameness… That shows you that the person trusts you if they’re willing to tell you something about themselves.” [P108]}
\end{center}

Developing common ground, through discovering shared interests and traits, was again seen as important to relationship transition. This was seen to develop further through repeated and shared experiences with others.

\begin{center}
\emph{"Only if we are open about discussing all of these things we figure out what our common interests are and if there are common interests of course then the conversation flows even better and by spending more and more time you become a friend.” [P103]}
\end{center}

\subsection{Purposes of Agent-Based Conversation}
\subsubsection{Transactional over Social Conversations}
There was a marked difference in the way that participants discussed having conversations with agents compared to conversations with other people. Conversations with agents were almost universally described in functional terms. Their status as an agent meant participants perceived a high barrier to reaching the more social and emotional connections seen in human conversation.

\begin{center}
\emph{"I would still think of a conversational agent as a tool.” [P109]}
\end{center}

\begin{center}
\emph{"Because it’s a machine you can’t make an emotional connection with it.” [P102]}
\end{center}

Participants identified that the concept of conversation may need to be reconsidered or defined around different parameters in agent-based interaction. Emulating human interaction was perceived as difficult, if not impossible, with users questioning whether this emulation was even desirable. 

\begin{center}
\emph{"So probably the conversation with a machine should be characterised by different aspects...like for example the clarity of the conversation.” [P104]}
\end{center}

\begin{center}
\emph{"What we see as conversation, I don’t think we can replicate it so I think there has to be new parameters for what a conversation is with a machine.” [P106]}
\end{center}

\subsection{Attributes of Conversation with Agents}
When asked to reflect on the attributes mentioned in human conversation, it became clear that these were conceptualised in markedly different terms when applied to human-agent conversation

\subsubsection{One Way Understanding and Personalisation over Common Ground}
Participants described building common ground and mutual understanding as integral to conversation with other people. Yet, there was a clear difference in the role that participants perceived common ground should play in conversational agent interaction. Some participants felt displeased when considering common ground operating in a similar way to human conversation. 

\begin{center}
\emph{"I’d be quite upset if I thought that I had common ground with a computer.” [P108]}
\end{center}

Instead, common ground was conceptualised as personalisation, where information would be remembered by the agent to tailor their experience. Participants noted that, over time this could create an illusion of common ground between a machine-like agent and user. 

\begin{center}
\emph{"I would find it very difficult to comprehend common ground with a machine. If you were to personalise your machine you might have the perception of common ground. For example, I like rugby so if my machine used rugby analogies explaining things to me I'd perceive it as having a common ground.” [P105]}
\end{center}

\begin{center}
\emph{"The more you interact with agents the more they learn about you so the more personalised it becomes...So machines keep learning from what you give as input to them.” [P103]}
\end{center}

Common ground was not perceived as co-constructed in agent dialogue. Participants felt that the system should lead this process, with the user making little effort to support this. 

\subsubsection{Functional Trustworthiness and Privacy}
Trustworthiness in human conversations was linked to sharing personal information and vulnerabilities to increase social bond. In agent conversations, trustworthiness was discussed exclusively in utilitarian terms. Responses related to security, privacy, and transparency over emotional trust.

\begin{center}
\emph{"Trustworthy when it comes to a machine is more about the security features that are built into it.” [P103]}
\end{center}

\begin{center}
\emph{"I think trust definitely in regards to data...you know is this machine recording our conversation? How is that information being used? How are you using my data? Who has access? I think that’s where trust might come in.” [P108]}
\end{center}

The definition of trustworthiness clearly lacked the emphasis on emotional trust seen when discussing human conversational characteristics. That said the conceptualisation of trust in this context may still act as a gateway to further interaction, in that issues of efficiency, reliability and security may be important for frequent long-term use of conversational agents.

\subsubsection{Accurate Listening}
Again, like other attributes, participants defined the role of the agent as a listener in more functional terms. Participants emphasised the need for the agent to understand them clearly and quickly, ideally without repeating themselves. Many of the comments focused around speech recognition performance. 

\begin{center}
\emph{"I think the voice recognition and not having to repeat yourself would be more of the receptive side of the machine...It’s not bettering your experience if you have to sit and repeatedly ask.” [P108]}
\end{center}

\begin{center}
\emph{"I suppose that’s number one. I want the agent to understand what I’m saying and be able to parse it properly.” [P111]}
\end{center}

\subsubsection{Humour as a Novelty Feature}
While humour was seen to scaffold human conversations, participants described humour as more of a novelty feature that can help make interactions with agents more interesting. Agents were seen to lack the ‘organic’ process of humour seen between people.

\begin{center}
\emph{"It [humour] makes the interaction interesting and you want to keep using these kind of devices which makes it fun for you to use it.” [P103]}
\end{center}

\begin{center}
\emph{"I don’t think you’re going to get Humour. I think it’s hit or miss. I know you can ask Siri to tell you jokes and sometimes they do and it’s usually on the dad jokes level. It’s not actually funny the way an organic conversation would be between two humans.” [P117]}
\end{center}

While the novelty of humour with agents was noted as a positive feature it was rarely described as necessary in the same way it was for human conversation.

\subsection{Relationship Building in Agent-Based Conversation}
\subsubsection{Becoming Friends with Conversational Agents}
As presented earlier, our analysis showed that current human-agent conversations serve primarily transactional purposes. From a relationship point of view, this echoes the transactional interactions people have with strangers or casual acquaintances. Conversation was identified as the cornerstone of becoming close friends in human interaction. A key question arose as to whether and how this can be accomplished with agents. In our data, participants could not overcome the functional purposes of human agent conversation and were resistant to the idea of becoming friends with something inherently machine-like. Conversational agents were consistently considered as tools and assistants available to serve and accommodate people. Motivations for building a different relationship with them were questioned.

\begin{center}
\emph{"I don't know why I would want my tool to be vulnerable, or be intimate, I think like you said it has-- like if I wanna put a nail in a table, I get a hammer. If I want to find out how to get to a particular place you'd put it into, Siri or something like that.” [P109]}
\end{center}

\begin{center}
\emph{"I dunno if you really want to sit and have a conversation that didn't require something out of the machine like the weather or turn this on or make an appointment for me or that kind of thing but... you're not gonna make friends with a machine so…” [P117]}
\end{center}

Participants dismissed friendships with machines as ‘un- normal’ or were reluctant to envisage having conversations with them in the same way as they would with their friends.

\begin{center}
\emph{"I mean I don't enjoy communicating with machines...when machines are like trying to become human that's just not ok...because. If you're talking to a machine to get their perspective on something, it'd be bit, un-normal.” [P112]}
\end{center}

\begin{center}
\emph{"It's not like you can chat with a chatbot about how you feel and why your morning sucked.” [P101]}
\end{center}

Conversing with agents was perceived as innately different to interacting with people, meaning friendship building could not be accommodated

\begin{center}
\emph{"But it's still not the same as with a human ‘cause the machine is gonna be like...running through algorithms to go what are we talking about-- keep in this range so they're not gonna interrupt halfway through go oh my god I forgot to tell you about this, or before I forget I have to tell you about that, that kind-- a machine is not gonna do that.” [P117]}
\end{center}

\subsubsection{Relationship Tensions}
Participants pointed to potential conflicts that can arise if people were to be friends with machines. Building a relationship with a conversational agent was seen to require significant time and effort together with a potential shift of the nature of friendship.

\begin{center}
\emph{"…we’re going to have to really reconfigure how we think about what friendship is for us for this to work…If that’s what people would want… because it’s not like having a pet…It will be very very interesting to see if people do start adopting…machines as companions in the home.” [P106]}
\end{center}

The prevailing perception of a master-servant relationship invokes key question in terms of how to reconcile ordering around a friend, or what would occur if an agent was to decline your request? 

\begin{center}
\emph{"It’s a strange thing, on one hand you’re sort of becoming a friend, but on the other you can order it to do what you want …or ask it what you want. In that sense it would be difficult to ever really think of it as a close acquaintance. Then it would have the right to say no to you. And then people throw those things out the window” [P113]}
\end{center}

A further tension involved the monetary value of agents for companies and concerns around how monetary incentives will operate in the context of a human-agent friendship.

\begin{center}
\emph{"It depends, to my idea, the main reason behind chatbots, is pretty much money - there is not really a conversation going on. It's more people want to create chatbots because they want to have a more flexible customer care - maybe less expensive. It's a good application of a nice and powerful technology but it's different from conversations itself, right?” [P101]}
\end{center}

\subsubsection{Potential Scenarios for a Human-Agent Relationship}
Despite initial reluctance participants did see opportunities for a human-agent relationship to be of value. Such scenarios involved people who are isolated such as elderly or struggling with mental health issues and could benefit from conversation.

\begin{center}
\emph{"Let’s say for example it’s a case…of anxiety. And you’re going, I have to go to this work thing, but I don’t want to go. And the machine will start telling you what the benefits are of going. Maybe it’s better for your career, maybe you’ll, get more chance…you’ll be seen. All this type of stuff.” [P113]}
\end{center}

\begin{center}
\emph{"I think that those are the only two, like task based and keeping the person company.” [P112]}
\end{center}

That said, when considering potential scenarios where conversational agent could be valuable, participants primarily described functional applications such as controlling home appliances, monitoring health, scheduling appointments or dealing with daily personal administrative tasks.

\begin{center}
\emph{"If you have something in the home that is able to guide and navigate people through their illness and teach them how to self-manage better. If it’s as simple as reminding them to take their medications… telling them about hospital appointments” [P104]}
\end{center}

\begin{center}
\emph{"Not nagging you but at least giving you prompts and giving you options of how you might actually conveniently do this...These things should definitely be linked into all that. They’re aware of your taxes as well holy God, nobody understands about taxes like. It’s all stuff you have to find out, that you know there’s this machine that’s able to take all of that admin stuff and even just organise it for you.” [P106]}
\end{center}

\section{Discussion}
Our study aimed to identify important conversational characteristics in human conversation and how these may vary when applied in conversations with agents. Conversational properties in human-human dialogue are well known and the current uses of speech agents are also well mapped. Yet people’s perceptions of conversing with machines (not just IPAs) are less understood. Through asking participants to reflect on the properties of conversation in both human and agent contexts, we aim to contrast the nature of human-human and human-agent conversation. Our findings show that compared to the perceived social and transactional purposes of human-human conversation, agent-based conversation was conceptualised in almost purely transactional terms.  Echoing seminal work in linguistics \cite{brown1987politeness,clark_using_1996,goffman_interaction_1967,jaworski2014silence, ziv_social_2010}, important characteristics of human-human conversation focused on mutual understanding and common ground, trust, active listenership, and humour. When reflecting on these characteristics in agent contexts people described them in highly functional terms. In conversations with agents, common ground development was viewed as a one-way process related to personalisation. Although this brings an illusion of mutual shared knowledge integral to common ground \cite{clark_using_1996, clark1992arenas}, participants clearly do not see this as being collaboratively built, negotiated and updated during agent dialogue, juxtaposing grounding processes in human conversation \cite{clark1991grounding}. Trust and listenership were defined in terms of system performance rather than important precursors for social bond, whilst humour was seen as a novelty rather than an integral component of conversation. 

Conversation was integral to developing long-term friendships between people. Conversely, participants fundamentally questioned the desire and ability to befriend or converse with an agent in the same way. They identified key conflicts in trying to do so (e.g. issues with master-servant dynamic, dissonance between monetary incentives and agent-friendship development). Core to many issues was the status asymmetry seen between users and conversational agents, where the agent is perceived as a tool or servant for the user. This creates a fundamental barrier to developing long-term relationships. That said, there were limited contexts where participants felt that conversation with agents would be beneficial, such as to support people who are isolated or struggling with mental health issues. 

\subsection{Current Perceptions as a Barrier to Conversational Agent Interaction}
Our data emphasised a fundamental mismatch in perceived status between agent and user where the agent was considered a user-controlled tool rather than a potential companion or social equal. This, in tandem with perceptions of functional ability \cite{moore_is_2017} may restrict the types of conversations that users perceive as appropriate or possible to have with an agent at present \cite{clarksocial_2018}. Critically, our findings suggest that social aspects of conversational interaction are currently absent from people’s perceptions of what conversational agents can and should be capable of performing. These expectations support the notion that an agent is a basic dialogue partner \cite{branigan2011role} lacking in humanlike conversational capabilities \cite{cowan_what_2017,luger_like_2016,porcheron_voice_2018}. Our participants personally expressed no desire to build bonds with conversational agents (although they expressed the view that other user group may find conversational features useful). The lack of enthusiasm for bonding may stem from the core belief that agents are poor dialogue partners that should be subservient to the user. It may also lie in the perception that there is no support for social dialogue in the current infrastructure for conversation \cite{dubberly2009modeling}, as more social and conversational dialogue is currently lacking in current VUIs \cite{porcheron_voice_2018}. This type of perception may be anchored by current experiences of IPAs, which are not designed to satisfy interpersonal goals, fuelling stereotypes around conversational agent abilities. The basic view of agent capability may be one of the most significant current barriers to users embracing or utilising conversational agents for social goals.

\subsection{Reframing the Concept of Conversation in Agent Interaction}
Our findings support the view that the paradigm of conversation with agents needs to be reframed \cite{porcheron_voice_2018, reeves_conversational_2017}. It is clear that participants in this paper categorised conversation with agents as almost exclusively task-oriented and transactional, echoing findings in other literature \cite{cowan_what_2017,luger_like_2016,porcheron_voice_2018}. Current commercial IPAs are not necessarily designed to deliver social conversation. However, there remain contexts where, because of the need to foster a long term and personal human-agent relationship, conversational agents may need orienting towards addressing interpersonal and social goals. Yet this may not need to emulate the form or outcomes of human conversation. Rather than simulating human conversational abilities in the hope of successful social conversation with users, our findings suggest that we need to treat human-agent interaction as a new genre of conversation, with its own rules, norms and expectations. As articulated in our data these may be more functional and utilitarian in nature, with little emphasis on the relational growth or emotional outcomes seen in human conversation. Our data suggests that the conversational content and structure present in interactions we have with strangers or acquaintances may be a good starting metaphor for social agent conversations. 

Indeed agent conversations may mirror more limited service-oriented or ‘front desk’ encounters between people. Talk is sequential \cite{sacks1987preferences}, and in limited encounters, conversation is often sequentially organised around adjacency pairs signifying requests for services followed by provisions of services \cite{sacks1978simplest}. They can also include openings to signifying the beginning of interactions (similar to IPA ‘wake words’) and closings to signify interactions ending. Human-human service encounters may also include an optional interrogative series, where more information is gathered \cite{kidwell2000common}. These features differ from modern speech technology which are often limited to isolated adjacency pairs without closings \cite{porcheron_voice_2018}. However, these conversational norms and indeed expectations of speech technology will likely be dynamic, shifting as long-term and multiple-turn agent use become more commonplace in contexts where they are designed to address social needs and not simply fulfill service requests. These conversational norms and expectations will likely be dynamic, shifting as long-term agent use becomes more commonplace in contexts where they are designed to address social needs.

The importance of context in shaping the content and norms of agent conversation cannot be overstated. For instance, what people deem appropriate to divulge or discuss conversationally with agents may differ markedly in private and public settings. Current users are unlikely to engage with IPAs in public \cite{cowan_what_2017,luger_like_2016}, noting social embarrassment and awkwardness \cite{cowan_what_2017, abdolrahmani2018siri}. Users have fewer concerns divulging private information when using VUIs in private compared to social spaces \cite{abdolrahmani2018siri, easwara2015privacy}. We identify a clear distinction between human and agent based conversation in terms of its perceived norms, rules and expectations. The context of interaction is no doubt likely to impact these and further work should look to explore this impact.

\subsection{Agent Conversation for Social Goals}
Our work questions the extent to which imbuing conversational attributes in agents will lead to similar benefits and relationships to those in human communication. However, in some contexts conversational capabilities may help facilitate interaction and use. For example, research in social robotics \cite{sabelli_conversational_2011} shows that elderly users may benefit from social capabilities in a system, leading them to disclose personal stories. A sensitivity to the context of interaction, an understanding of the type of conversation required, and the purposes behind the interaction are integral for this to succeed. As suggested in our data, conversation with agents may not be the best approach to use to develop bond. This could be achieved through exploiting other modalities, especially in more embodied systems (e.g.\cite{goffman_interaction_1967}). Through gesture and facial expressions cues (e.g. \cite{breazeal_emotion_2003,bruce_role_2002,kuno_museum_2007}) embodiment may even help support speech-based conversation. 

\subsection{Limitations and Future Work}
This paper identifies common characteristics people view as important in conversation and how these differ when applied to conversations with agents. Participants saw interpersonal relationships as immaterial when engaging with conversational agents. These views may come from a lack of familiarity with agents that are designed to engage in conversation. In our data it is clear that participant views were grounded by existing IPA experiences and the types of interactions those agents facilitate (e.g. short-term, sporadic and transactional). Future research should explore ways to get users to focus on envisioning longer, sequential interactions where conversational capabilities are more sophisticated \cite{reeves_envisioning_2012,rodden_at_2013}. Although we did not get participants to interact with conversational agents during the research, getting users to engage with a (real or simulated) conversational agent, may act as a useful trigger to support generation of characteristics and scenarios. These approaches could be particularly valuable for the design of future conversational agents, identifying issues that may arise in their implementation. Future research would also benefit from recruiting participants who are less familiar with technology. Most participants in this study described themselves as intermediate, advanced or expert users of technology. People who are further away from understanding how IPAs and similar technology works may have different opinions towards agent-based conversations. Recent studies have indicated that certain users groups tend to anthropomorphise their devices, particularly older adults and children (e.g.\cite{purington2017alexa, yarosh2018children}). Users such as these may perceive conversation with machines differently to the sample population in this paper.

\section{Conclusion}
Conversations play an essential role in building and maintaining relationships with other people. However, this is not seen as important or even desirable in most current scenarios with conversational agents. While many concepts of good human-human conversation are discussed as important in human-agent conversations, how they are described is markedly different. Participants describe mutual understanding and common ground, trust, active listenership, and humour as crucial social features in human conversations. In agent conversations, these are described almost exclusively in transactional and utilitarian terms. Our findings suggest there may be a limit to the extent interactions with agents can mirror those with other people. These seem to be influenced in particular by existing agents whose interactions stray from common definitions of conversation. However, there may be specific application areas where conversation may be appropriate if not essential between humans and agents, particularly in areas such as healthcare and wellbeing, where the nuances of contexts and demographics need to be considered. This paper suggests that conversational agents can be inspired by human-human conversations but do not necessarily need to mimic it. Instead, human-agent conversations may need to be seen as a new genre of interaction.  

\begin{acks}
\grantsponsor{<IRC1>}{<Irish Research Council>}{<https://research.ie>} \grantnum[<https://app.dimensions.ai/details/grant/grant.7516751>]{<IRC1>}{<R17339>}
\end{acks}

\balance
\bibliographystyle{ACM-Reference-Format}
\bibliography{chi2019}

\end{document}